# Complex Aperture Networks

H.O. Ghaffari[1], M. Sharifzadeh[2] and E. Evgin[1]

[1]*Department of Civil Engineering, Faculty of Engineering, University of Ottawa, Canada*

[2]*Department of Mining & Metallurgical Engineering, Amirkabir University of Technology, Tehran, Iran*

**Abstract:** A complex network approach is proposed to study the shear behavior of a rough rock joint. Similarities between aperture profiles are established and a general network in two directions (in parallel and perpendicular to the shear direction) is constructed. Evaluation of this newly formed network shows that the degree distribution of the network, after a transition stage falls into a quasi stable state which is roughly obeying a Gaussian distribution. In addition, the growth of the clustering coefficient and the number of edges are approximately scaled with the development of shear strength and hydraulic conductivity, which can be utilized to estimate, shear distribution over asperities. Furthermore, we characterize the contact profiles using the same approach. Despite the former case, the later networks are following a growing network mode.

## 1. Introduction

A thorough understanding of the behavior of rock joints or fault surfaces is the first requirement in the study of abrupt motion, seismicity or flow patterns in geomaterials. A quantitative representation of the complexities involved in either the mechanical or hydro-mechanical characteristics of rock joints is an effective approach in the solution of large scale problems [1-3],[10-17]. A well-known model that captures the statistical features of the stick-slip behavior of joints during the fracture evolution is the Olami-Feder-Christensen (OFC) model. This model successfully gives the general evolution of interacting elements of a rough fracture under slow stress increments [21]. The interaction between the elements of a rock joint determines its behavior. However, one cannot predict the collective behavior of the whole by merely extrapolating from the treatment of its units [5-6]. One possible approach in describing the interconnected systems using physical tools as a major reductionism is to simplify the interactions between the elements. The complexity reduction is a rescue pathway to obtaining a solution. However, it is an approximate solution to a collective behavior of particles having swing states, complicated structures, and diversity of relations among elements [9, 19]. On the other hand, dissection of phenomena under investigation into a list of interacting units - which are the building blocks (granules) of a system and are connected by pair-wise connections, can reveal the presence of complex networks, which provide a mathematical framework for the analysis of wide range of complex systems. Picturing, modeling and evaluation in a simple and intuitive way are some of the distinguished features of complex networks [4-5]. Related to geosciences fields, Complex earthquake networks, climate networks, volcanic networks, river networks, which require large scale measurements, have been taken into account. In small scales, topological complexity has been evaluated in relation to geosciences fields such as the gradation of soil particles, fracture networks, aperture of fractures, and the mechanical behavior of granular materials [22-35].

Related to our present work, the models for the stick-slip behavior of fractures, spreading networks over space and time, the "events" (such as an abrupt change of asperities) and the OFC model are important developments to mention [21,24,36]. In our earlier studies, we captured the structural complexity of rough surfaces and scaling with the properly constructed networks [17,37]. In addition, we used a network approach on aperture profiles which was constructed by employing two measure criteria: "Euclidean, correlation and Chebishev". The current work will cover new aspects of Euclidean measure of an aperture-network such as the resolution effect, k-c space of complex aperture networks and stair-like profiles (contact profiles). We will show the scaling of development of the frictional forces with the attributes of the proper networks, which will give approximately the evolution of shear stresses acting on the profiles. This result coincides with the prediction of the OFC model on the stress distribution [38].This paper presents a complex network approach over aperture profiles where the mechanical and hydraulic properties of a rock joint are compared with the network properties. The next section covers the details of the method and the general aspects of network measures. After that the results of experiments on a rock joint and the arranged complex networks are shown.

## 2. Network on Apertures

To set up a network on apertures of a joint under a certain amount of normal stress, two different strategies may be followed: (1) In the first case, each surface of the joint is replaced by a network. This step is based on a certain measure over elements of each surface. Subsequently, the interactions of these two networks are evaluated. Here, the objective is to find out the relations between the characteristics of the networks and the overall rock joint properties. Achieving this objective within a suitable error level is time consuming, especially when we are dealing with huge data sets obtained by scanning the surfaces. (2) In the second approach, a network is constructed on the enclosed space between the two surfaces, i.e. complex aperture network. In the construction of the network over the openings, we consider strips of elements lying beside each other. Each strip is called a profile. In the aperture patterns, we consider each aperture profile[1] as a node. The network is constructed using these profiles representing the two surfaces. This paper follows the second approach. We assume that the upper surface is moving and the lower surface is fixed. The number of X-profiles (nodes) slowly decreases as the shear displacement increases in the x-axis direction. The labeling of the nodes is done based on the common area between the moving surface and the fixed one (Fig.1). To draw an edge between two nodes [18], a relationship (or in general relationship matrix) should be defined. By assuming that there are some hidden metrics between two nodes, the similarity between the nodes is captured. As the simplest metric, Euclidean distance, we will have:

$$d_{Euc.} = \sqrt{\sum_{p,q=1}^{n_p} (p(x_1, x_2, ..., x_n) - q(x_1, x_2, ..., x_n))^2}, \qquad (1)$$

1- Definitions: X-profiles: aperture profiles parallel to the Y-axis direction (perpendicular to the shear direction) and Y-profiles are parallel to the shear direction.

where $p$, $q$ are the $i^{th}$ profiles ,through $n_p$ profiles. $x_n$ is the height (aperture) of $n^{th}$ pixel (cell) in a profile. When $d \le \xi$, an edge between two nodes is created. The sensitivity of the network parameters with regard to aperture networks has been investigated in [17]. Here, we use $\xi$ as 5-20 percent of the maximum $d$. This step of the analysis is based on the granularity of the collected information [17]. Choosing a constant value for $\xi$ is associated with the current accuracy of the data accumulation in which, after the maximum threshold, the system (the aperture evolution) loses its dominant order [37,39].

One of the main attributes of networks is the clustering coefficient. The clustering coefficient shows the collaboration (loops and then circulation of information) among the connected nodes. Assume the $i^{th}$ node to have $k_i$ neighboring nodes. There can exist at most $k_i(k_i - 1)/2$ edges between the neighbors (local complete graph). If we define $c_i$ as the ratio [7]:

$$c_i = \frac{\text{actual number of edges between the neighbors of the } i^{th} \text{ node}}{k_i(k_i - 1)/2} \tag{2}$$

then, the clustering coefficient is given by the average of $c_i$ over all the nodes in the network:

$$C = \frac{1}{N}\sum_{i=1}^{N} c_i . \tag{3}$$

For $k_i \le 1$, we define $C = 0$. The closer $C$ is to one the larger is the interconnectedness of the network. The clustering coefficient can also give an indication of the number of triangles around each node. Therefore, it can be considered as a measure to circulation of information over structures. Sometimes, such information flow brings tortuosity and somehow, causes trapping and slowness in information propagation through nodes. Another property which can be used in characterization is the connectivity distribution (or degree distribution), $P(k)$, which is the probability of finding nodes with $k$ edges in a network. In large networks, there will always be some fluctuations in the degree distribution. Large fluctuations from the average value ($<k>$) refers to the highly heterogeneous networks while homogeneous networks display low fluctuations. There are different types of network models which have been investigated extensively. For example, the random networks of Erdős–Rényi [19], the small-world networks of Watts-Strogatz [4,7-8], and the scale-free networks of Albert-Barabasi [5, 6] are the most well known network models. Regular networks have a high clustering coefficient ($C \approx 3/4$) and a long average path length. Random networks (its construction is based on random connections of nodes) have a low clustering coefficient and the shortest possible average path length. The small-world networks have high clustering coefficients and small (much smaller than the regular ones) average path lengths.

## 3. Results

In this part, we focus on the experimental results and analysis of the hydro-mechanical properties of the rock joint using the approach described in the previous section. It is recognized that the network anatomy is very important in the characterization of the system's out-put (because the structure always affects the function). Our objective is to determine the possible relations between the constructed network properties and the measured mechanical / hydro-mechanical properties of a rock joint. To investigate the small world properties of the rock joint, the results of several laboratory experiments were used. In the experiments, the joint geometry consisting of two surfaces and the aperture between these two surfaces were measured. The shear and flow tests were performed later on. The rock was granite with a unit weight of 25.9 kN/m$^3$ and a uniaxial compressive strength of 172 MPa. An artificial rock joint was made at the mid height of the specimen by splitting. A special joint creating apparatus, is used for this purpose, which has two horizontal jacks and a vertical jack [1,16]. The sides of the sample are cut down after creating the joint. The final size of the sample was 180 mm in length, 100 mm in width, and 80 mm in height. A virtual mesh having a square element size of 0.2 mm was generated on each surface and the aperture height at each position was measured by a laser scanner. The details of the procedure can be found in [1,16-17]. Figure 2 shows the relevant information related to the measured parameters. The samples were subjected to a constant normal stress of 1, 3 and 5 MPa-see Figure 3a) in different experiments while the variation in surface heights was recorded. Figure1 shows the shear strength evolution for different normal stress values. In this study, we focus on the patterns, obtained from the test with a 3 MPa normal stress.

In the experiments, a special hydraulic testing unit is used to allow linear flow (parallel to the shear direction) intermittently with the shear displacements of the rock joint. The relation between the hydraulic conductivity and the joint aperture is given by Darcy's law:

$$Q = K_h A i, \tag{4}$$

where $Q$, $A$, $i$, $K_h$ are the amount of volumetric flow, area and hydraulic gradient, respectively. In the use of this equation, it is assumed that the joint is consisted of two smooth parallel plates. The flow rate and the hydraulic conductivity can be written as follows:

$$Q = \frac{g e_h^2}{12\upsilon}(w e_h) i, \tag{5}$$

$$K_h = \frac{g e_h^2}{12\upsilon}, \tag{6}$$

where $g$, $e_h$, $\upsilon$ and $w$ are the acceleration of gravity ($m/s^2$), hydraulic aperture, kinematic viscosity of fluid and the width of the specimen, respectively.

The analysis is carried out for the rock joint which is under a constant normal stress and is subjected to translational shear displacement with a constant rate of movement (Fig.1). By setting a threshold value $d \leq 5$ in Euclidian distance (Eq. (1)) and building a complex network (Fig.1a) on the X-profiles, gradual changes of the adjacency matrix form of the appeared networks are obtained. The results demonstrate that, after a phase transition step, the patterns of similarities are restricted to the adjacency of each profile [37]. Except for the boundary profiles, the influence distance on each side of a profile changes from 2 to 20 pixels (0.4-4 *mm*).

A similar procedure is ensued at the Y-profile network. It is noticed that the influence distance is different from one profile to another one. In other words, in Y-profiles, the influence distance is lower than that of the X-profiles. Also, after a transition stage (near 2*mm*), by decreasing the mesh resolution some insignificant features of the variation of clustering coefficient (scaled with the hydro-mechanical properties) are neglected (Fig.4).

A careful consideration of the changes taking place in the number of edges and the local clustering coefficient during shearing reveals some important properties of the network.

Gradual changes of connectivity frequency, either in X or Y profiles, display a similar transitional behavior: a transformation from a nearly single value function to a semi-stable Gaussian distribution [17, 37]. It can be shown that the degree of nodes changes and a transition to a semi-stable stage occurs with a larger convergence rate. In other words, similarities in Y-profiles occur faster than the similarities in X-profiles. The complementary results obtained from the calculation of clustering coefficients emphasize that the inverse values of $C$ - in X and Y-profiles- produce similar patterns as the ones observed in the evolution of the frictional forces and hydraulic conductivity (Fig.3). The similarities between the plot of $1/C_Y$ and the changes of the hydraulic conductivity (rather than the dilation behavior of the joint) proves that the progressively changing positions of the upper and lower surfaces of the joint (or increment of mean aperture) is not the only important parameter in the fluid flow. If we consider the clustering coefficient as an indication of the circulation and longer transformation of information flow, the inverse of that may be assumed as a sign of the flow conductivity or easy-path flow out of the traps. It is a noteworthy observation that the reduction in joint roughness and the total ensemble of similar asperities are the main possible agents in the flow behavior as well as the shear behavior of the joint. Particularly, at a step when interlocking of (SD=1) asperities and simultaneously decreasing of flow pathways take place, the maximum energy must be used to force fluid to flow through the profiles. It is noticed that the sensitivity of the clustering coefficients to the scaling is very high. For instance, with an increase of 5 to 10 times of virtual meshes (1-2 mm), the details of the $C$ fluctuations are omitted, although, the general form of variations is preserved as it is seen in the experimental outputs in particular for shear displacements $0 \leq SD \leq 2$ (Fig.4). Anisotropy in the rock joint behavior related to the shear displacement or fluid flow can be noticed by viewing the plots of $C_Y / C_X$ or $K_Y / K_X$ variations as a function of shear displacements (Fig.5). At the quasi-residual stage, these rates are inclined at 0.95 and 0.4 and at SD=3 mm, $C_Y / C_X$ takes highest value while in SD=2, $K_Y / K_X$ reaches its lowest value. It should also be noted that in $3 \leq SD \leq 4$, $\left(\frac{d(C_Y / C_X)}{d(SD)}\right) \prec 0$ and at the same time $\left(\frac{d(K_Y / K_X)}{d(SD)}\right) \succ 0$.

The appearance of nearly the same patterns in the evolution of network characteristics related to the mechanical or hydraulic properties of the rock joint can be seen by using some other metrics (such as correlations [36]). Let us assume that (see Figure 3) $\frac{1}{C_Y} \sim K_h^y$ $and$ $\frac{1}{C_X} \sim K_h^x$ .$Then$ $\frac{C_Y}{C_X} \sim \frac{K_h^x}{K_h^y}$. Now, let us focus on two different cases related to information flow and the comfortability (ease) of the flow: at around SD=3 $\frac{C_Y}{C_X} > 1 \rightarrow K_h^x > K_h^y$ then the fluid motion in perpendicular direction to the shear vector is easier than Y-profiles (parallel to shear). In other words, a rapid transformation of information is perpendicular to the shear vector. However, in other cases ($SD \neq 3$) $\frac{C_Y}{C_X} < 1 \rightarrow K_h^x < K_h^y$ and the flow parallel to the shear direction is easier and faster. The similar reasoning can be followed by employing the scaling of shear strength and network characteristic(s).

With introducing a new space associated with $k-c$ and assuming $1/<c_x^i> \propto \tau^i ; 1/<c_y^j> \propto K_h$, the variation of profiles can be mapped to a proper space, namely, $k_{x,y}^{i,j} - 1/c_{x,y}^{i,j} \left(i \in X-Profiles, j \in Y-profiles\right)$ (Fig.6). The patterns of the clusters in this space can be compared with the variations of the mechanical or hydro-mechanical properties of a given single joint (Fig.6a, b). A faster adaptation in Y-profiles is the distinguished feature of $k_y^j - 1/c_y^j$ space rather than $k_x^i - 1/c_x^i$ space in which there is a discerned discontinuity between two large apparent clusters (at $SD=2 \rightarrow SD=4$). Using an intelligent clustering method -based on competition in the hidden layer of a neural network model and under self organizing feature map (SOM) [22] - the appeared space $k_y^j - c_y^j$ with an elementary topology of $20 \times 20$ (in the second layer of neural network and over 500 epochs), the dominant structures of this space are identified. The clustered space can be granulated in three categories (Fig.6c): (1) increasing of $c_y^j$ - Decreasing $k_y^j$ ( increasing $SD$ - decreasing $K_h$ ), (2) increasing $c_y^j$ - increasing $k_y^j$ ( increasing $SD$ - increasing $K_h$ ), and (3) increasing $c_y^j$ - slightly change $k_y^j$ ( increasing $SD$ - slightly change of $K_h$ ). In Figure 6.d we have shown the evolution path of the profiles over the defined space which clearly exhibits the details of the synchronization of profiles. Such space as a way to show a synchronization pathway has been used in correlation measure over profiles of the apertures [36]

The approximate scaling of the inverse clustering coefficient in X-profiles with the shear stress evolution can be used as the criterion to estimate the distribution of frictional forces over the perpendicular profiles through the fracture evolution (Fig.7). The evolution of shear stress distribution demonstrates that at initial stages, the frictional forces of profiles cover a wide range of Gaussian like distribution, while at quasi-stable stages the concentration of stresses follows a similar one with smaller scope (standard deviation). It is a noteworthy observation that the clustering coefficient distribution can mimic approximately the results of OFC model (by using proper parameters) [38]. In a similar way, based upon $c_y^j$ - related to $K_h$ - reveals that at initial stages of the shear displacement the overall pattern of distribution is formed and almost in all other steps a significant disruption (transition) is not recognized (Fig7.b).

Using a different approach, we constructed a network based on contact profiles. In other words, the role of contact zones is separated from non-contact zones. Similar to the previous measure, the Euclidean distance is utilized while non-contact zones and contact pixels are transformed into the 0 and 1, respectively. Construction of the networks exhibits a new type of contact profiles evolution: Growing networks either in X or Y profiles (Fig.8 and Fig.9). At the initial stages of the evolution, neither active nodes nor edges are evident. This is due to the high fluctuations of contact profiles which are in the stair like shapes. The variation of connectivity degree distribution - especially for X profiles - (Figs. 8a and 9a) shows a transition from an exponential to a different type of distribution. This implies that at the quasi- residual stage the contact zones are concentrated in the specific clusters. Nearly same patterns were obtained using the correlation measure and the two-point correlation concept over either the parallel or perpendicular profiles [36]. Increments of displacement induce a lower instability of contact profiles (growing of non-contact areas) which force an increase in the similarity (Fig.8c and 9c). However, for instance in X-profiles, after a certain step (SD=6) development of new sites stops while the growth of edges continues. In addition to this event at $\frac{\partial N}{\partial t} = 0$ (after SD=5 – compare with Figs.3a and 3c), we have (at the same scale) $\frac{\partial C}{\partial t} < \frac{\partial K}{\partial t}$ which confirms the promotion of connectivity with other nodes. This interval is followed by a reduction in roughness which represents (even though the percentage of contacts goes to nearly a constant value) a continuously increasing similarity of contact profiles (a coherent variation of contact patterns).In addition, we proposed an algorithm based on the general preferential attachment-detachment concept [17] using experimental observations, detailed measurements over profiles, and simulations of water flow through the rock joint,

## Conclusions

Understanding the mechanism of fracture shear behavior is not presented clearly so far. It is due to lack of capability to measure and observe the changes of aperture and contact pattern elements. On the other hand, understanding the evolution of fracture geometry or contact and aperture cells pattern requires more research. Complex aperture network is a new and appropriate tool to predict the aperture and contact cells pattern evolution during shear. Therefore complex aperture network is used to study the evolution of fracture geometry during the shear. Modeling the mechanical, hydraulic and hydromechanical behavior of rock fractures based on rock joint geometry could be possible using this kind of research. Results of complex aperture network modeling shows coincidence with the obtained experimental results.

By considering the complicated behavior of a rough rock joint which is subjected to shear displacements, a network associated with a similarity measure was constructed at two different directions. The network properties provided a suitable correspondence with the empirically obtained mechanical and hydro-mechanical characteristics of the joint. In addition, to highlight the functionality of contacts, a separate network based on binary profiles is utilized. Analysis of the results through network parameters shows a transition of contact clusters density related to shear displacement. At initial displacements, formation of similar contact profiles is significant but after a certain amount of displacement similar contact profiles with occupation in three district patterns are observed. These three different patterns are separately scaled with a power law while the power was changing. There are clear similarity between fracture shear stress – shear displacement , hydraulic conductivity – shear displacement obtained from experimental test and the results of inverse clustering coefficient ($1/C_x$) and ($1/C_y$), clearly shows that Complex aperture network could successfully model the joint behavior. After calibration of complex aperture network it is possible to predict joint mechanical, hydraulic and hydromechanical properties using joint geometry, which were scientific goals for several years and researchers. Understanding the joint or fracture geometry evolution during shear has key role to understand the fracture mechanical, hydraulic and hydromechanical and other properties.

The following points are the main results of our work:

1-Scaling of complexity structures over apertures of a rock joint with hydro-mechanical properties

2- Analysis of anisotropy in hydraulic conductivity analysis using networks characteristics

3-Inferencing the hydro-mechanical properties evolution using the *k-c* space of aperture

network

4-Estimation of shear forces and conductivity distribution through the profiles.

# Figures

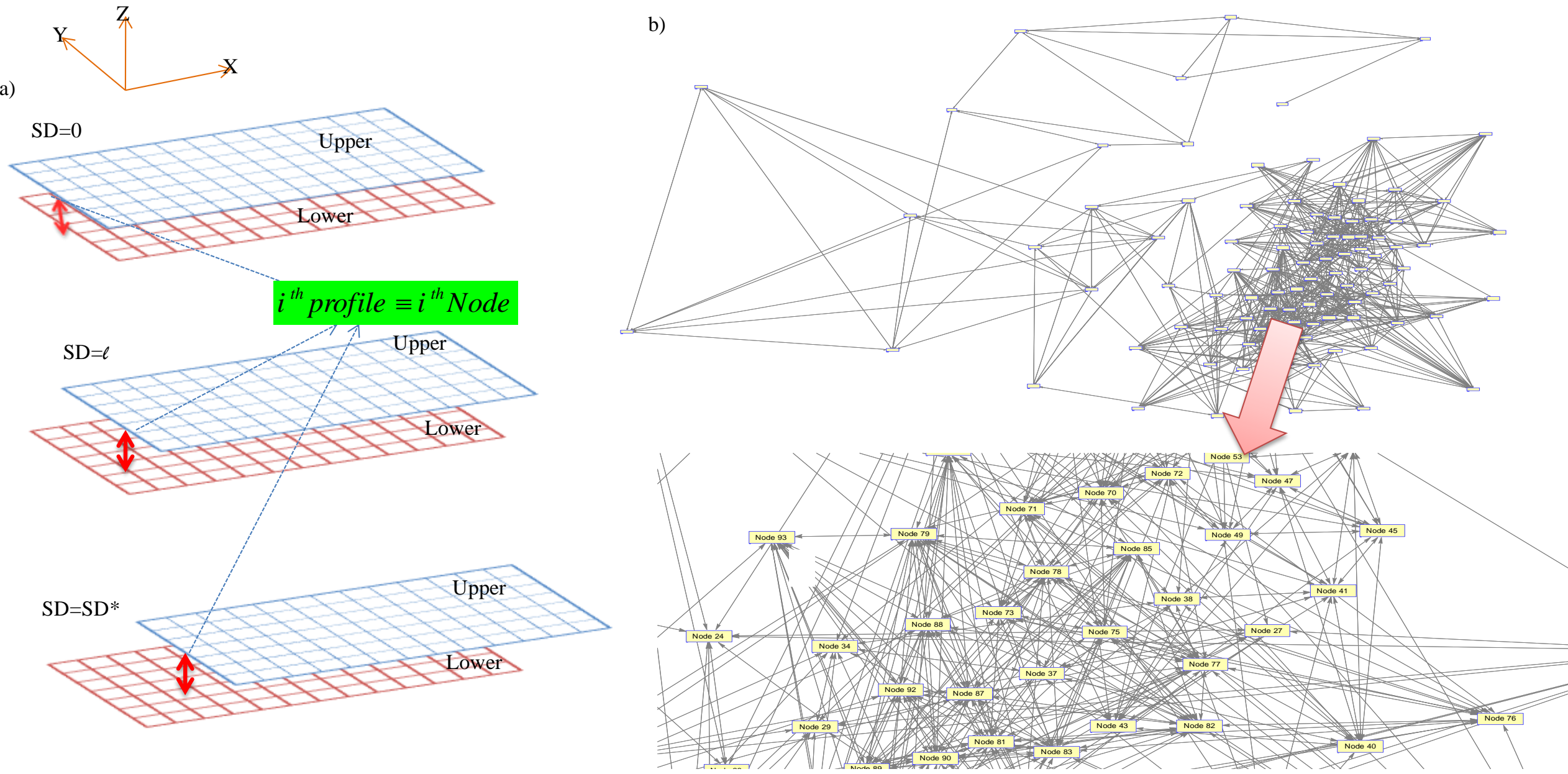

**Figure 1.** a) Construction of a complex network based on the measured aperture profiles (the node references correspond with the upper surface of the joint), b) A small section of the created network (only the first 100 nodes out of 801 nodes are shown) at 20 mm shear displacement (SD) on X-profiles (perpendicular to shear direction).

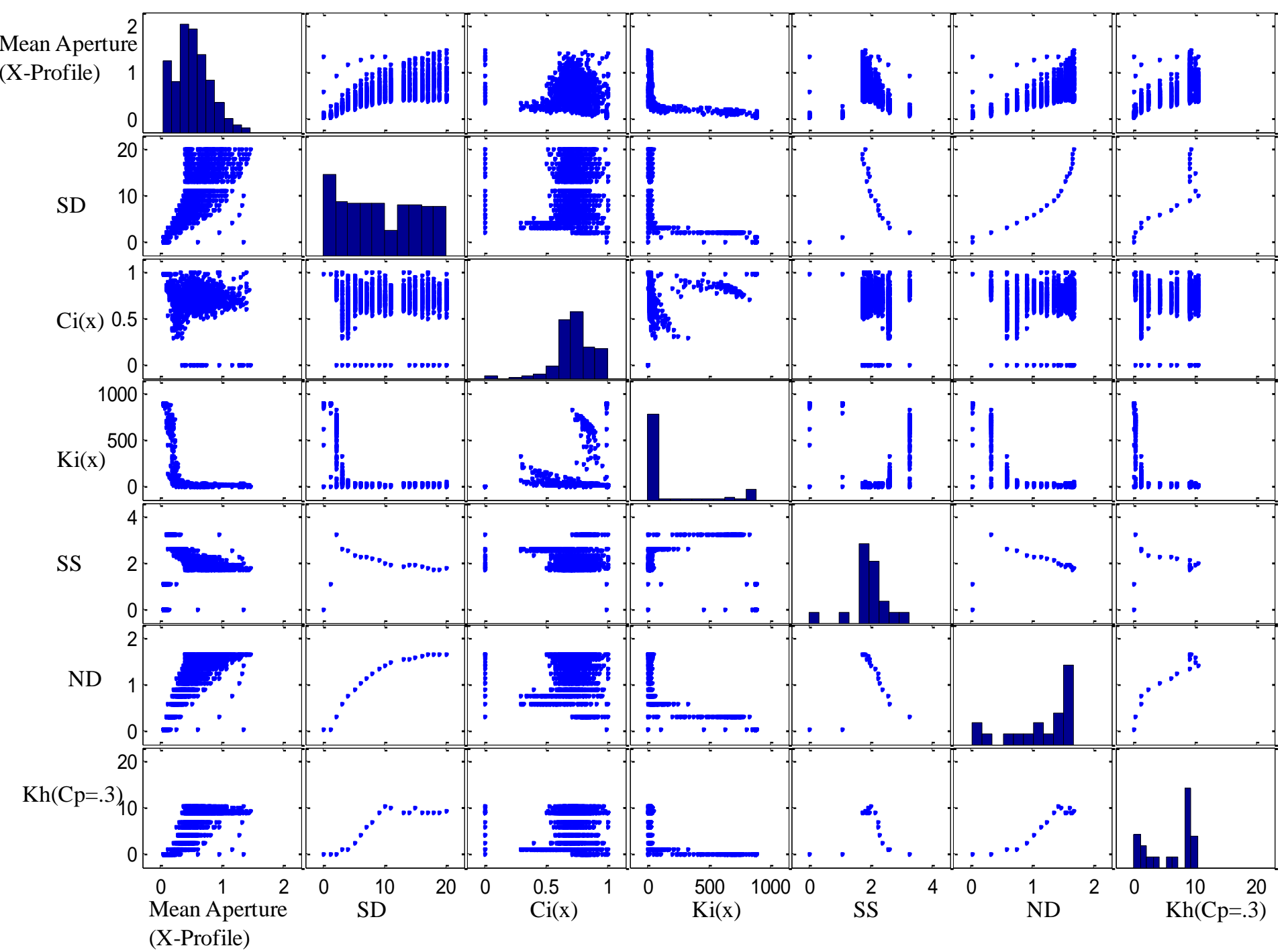

**Figure 2. The gathered** Information from experimental and network analysis (scatter graph of the measured information/data) during shearing of a rough joint: Shear Displacement (SD in mm); Clustering coefficient of each node in X-profiles ($c_i(x)$); Degree of node ($k_i(x)$); Shear Stress (SS in MPa); Normal Displacement (dilatancy-ND in mm) and Hydraulic Conductivity ( $K_h$-cm/s). The diagonal plots show the distribution of attributes (without frequency values in this case). Each sub-plot shows two-variable states of the recorded data set through successive shear displacements. For example SD-Mean aperture (of each profile) shows growth of openings with the displacements; $c_i(x)$-SD and $k_i(x)$-SD show abrupt drop around peak point (see Figure 3). The classical behavior of evolution of frictional forces and dilatancy of the joint are shown in SS-SD and ND-SD plots. Spectrum of the network through cumulative measured information is in $c_i(x)$- $k_i(x)$( see Figure 6 for more details).

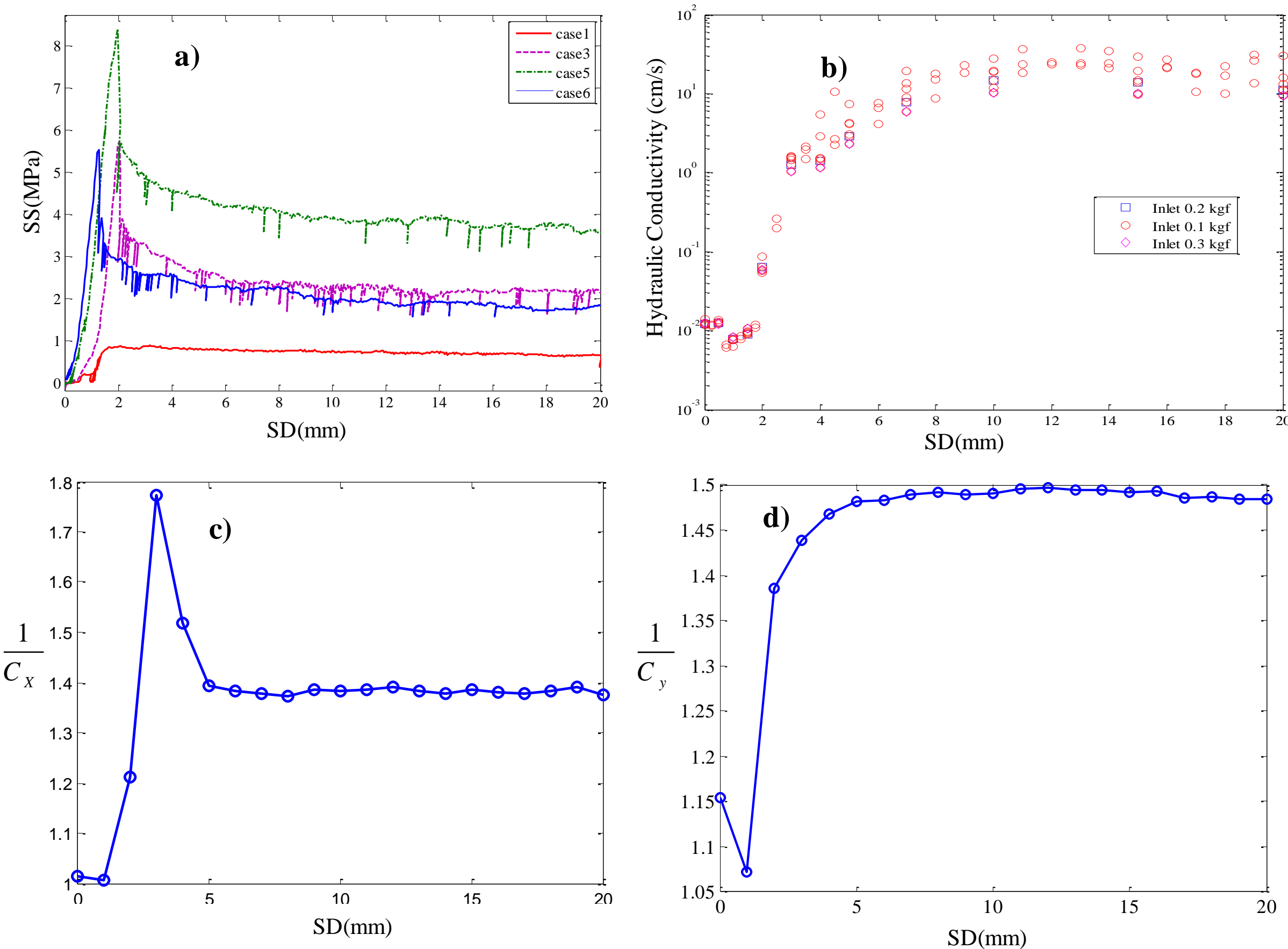

**Figure 3**. a) Experimental results of the shear stress versus shear displacement (case 1, 3,5 and 6 are corresponding to different situations of experimental procedure ,which are depends on upper and lower boxes position and different normal stresses during the test [1,16]- here, we use the case 3 with 3MPa normal stress ), b) (Experimental results) Hydraulic conductivity versus shear displacement associated with 3 different cases, c) & d) (calculated by network model) Inverse of Clustering Coefficients versus shear displacement on the X and Y profiles, respectively.

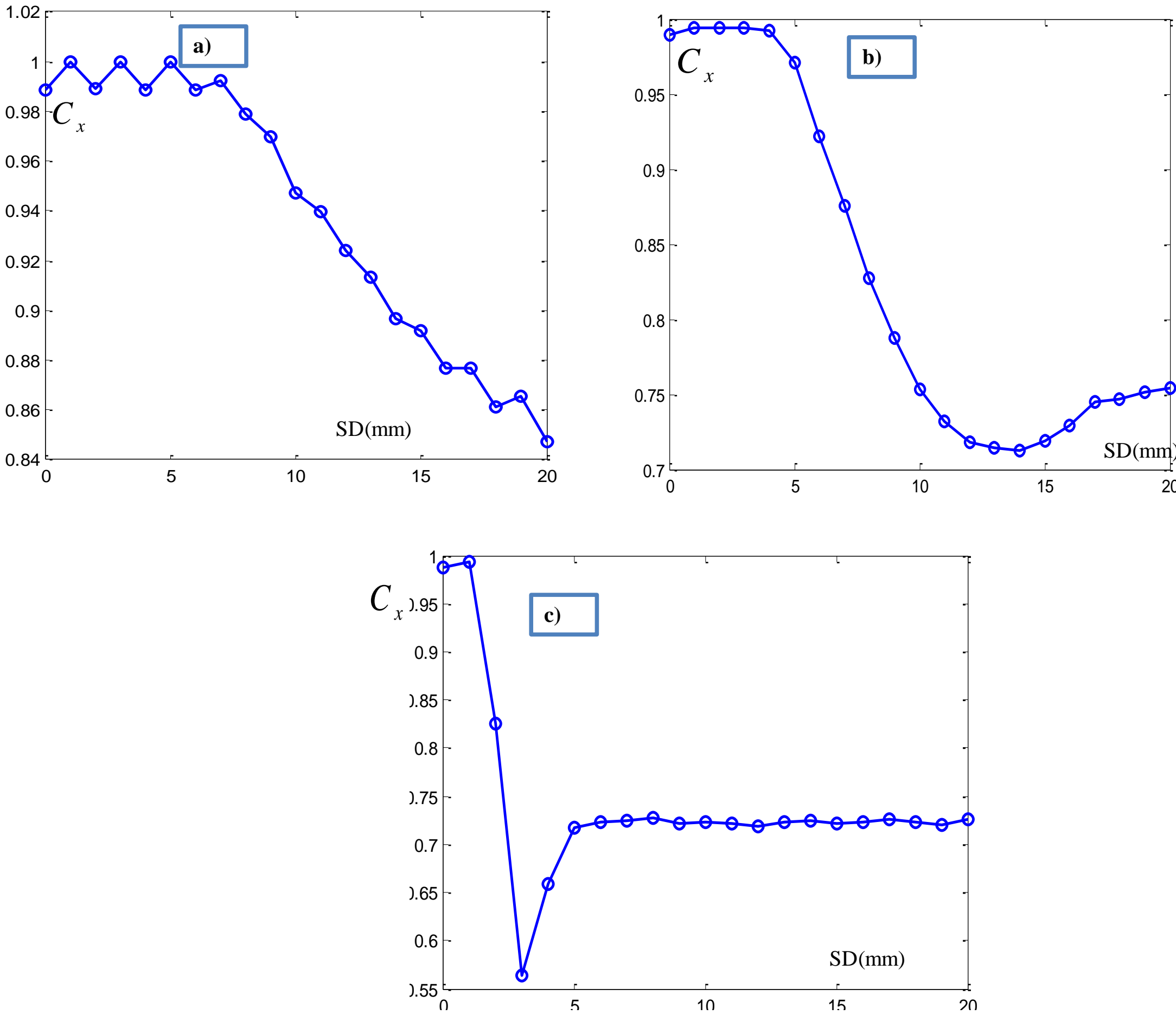

**Figure 4**. The resolution effect on the Clustering Coefficient evolution calculated from the network model: when the profiles are selected within a)10, b) 5 and c) 1 pixels interval. High resolution of the measured apertures discloses much more realistic picture of the structural complexity of fracture planes.

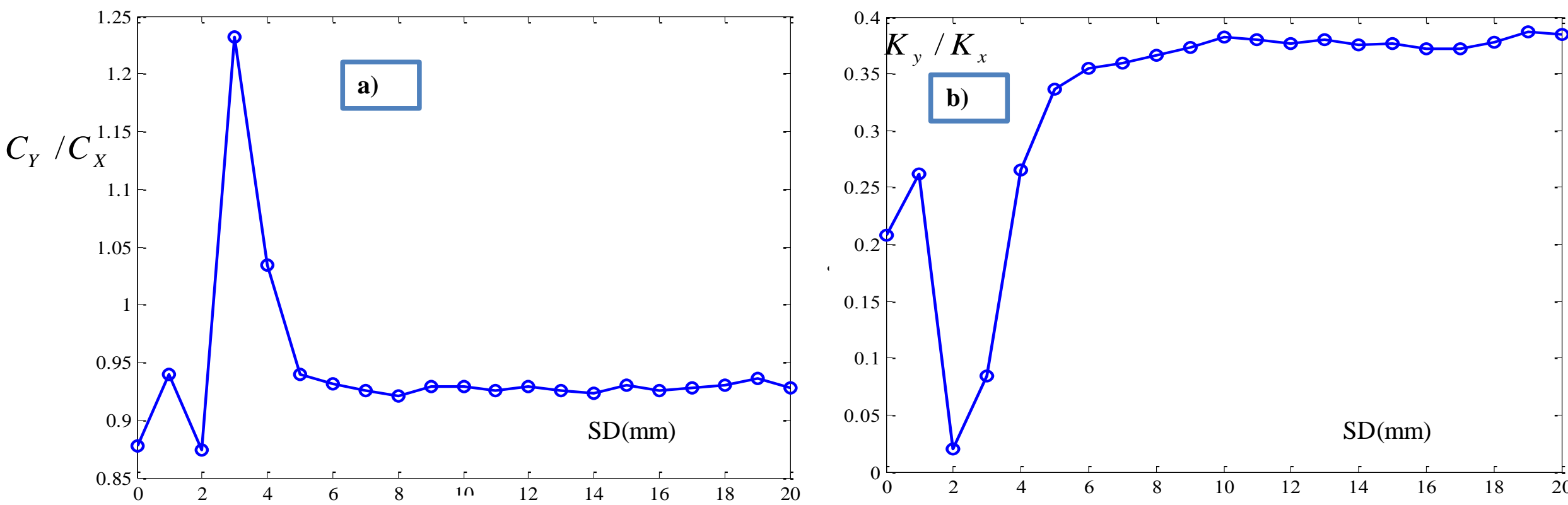

**Figure 5**. Anisotropy in evolution of parallel and perpendicular networks (calculated based on the network model) a) Variation of $C_y / C_x$ (clustering coefficients) with shear displacement SD; b) Variation of $K_y / K_x$ (nodes degrees) with SD during shearing.

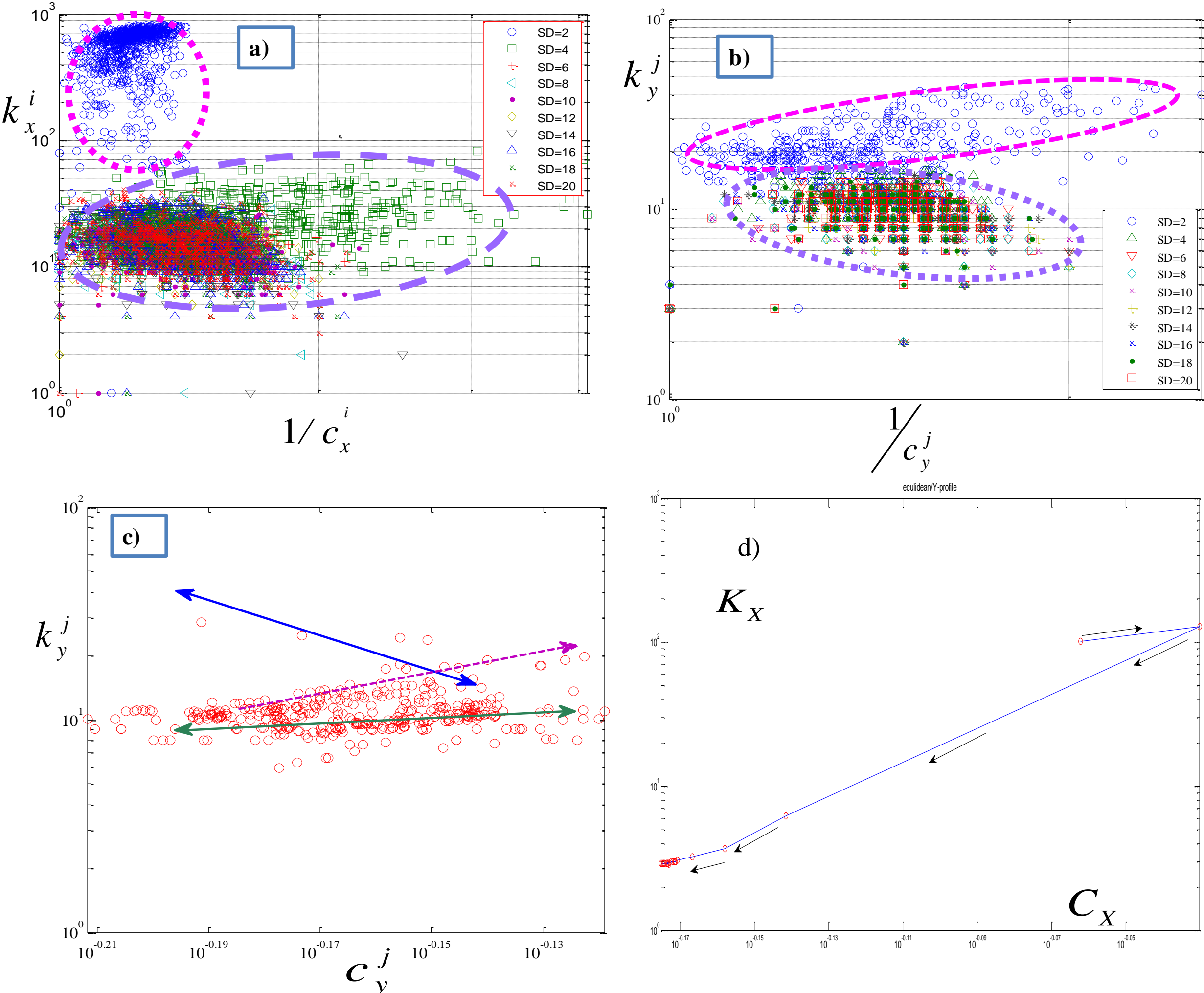

**Figure 6**. a) Evolving $k_x^i - 1/c_x^i$ state space; b) Changes in $k_y^j - 1/c_y^j$ relation during shear displacements (Oval shapes indicated by dashed lines appear to be the main clusters); c) Clustering of $k_y^j - c_y^j$ space using SOM (neural network) with initial topology $n_x = 20, n_y = 20$ in competition layer after 500 epochs training (d) Path evolution of the joint in the $K_X$-$C_X$ space. All of the graphs are the results of the calculated networks parameters over observed experimental data set.

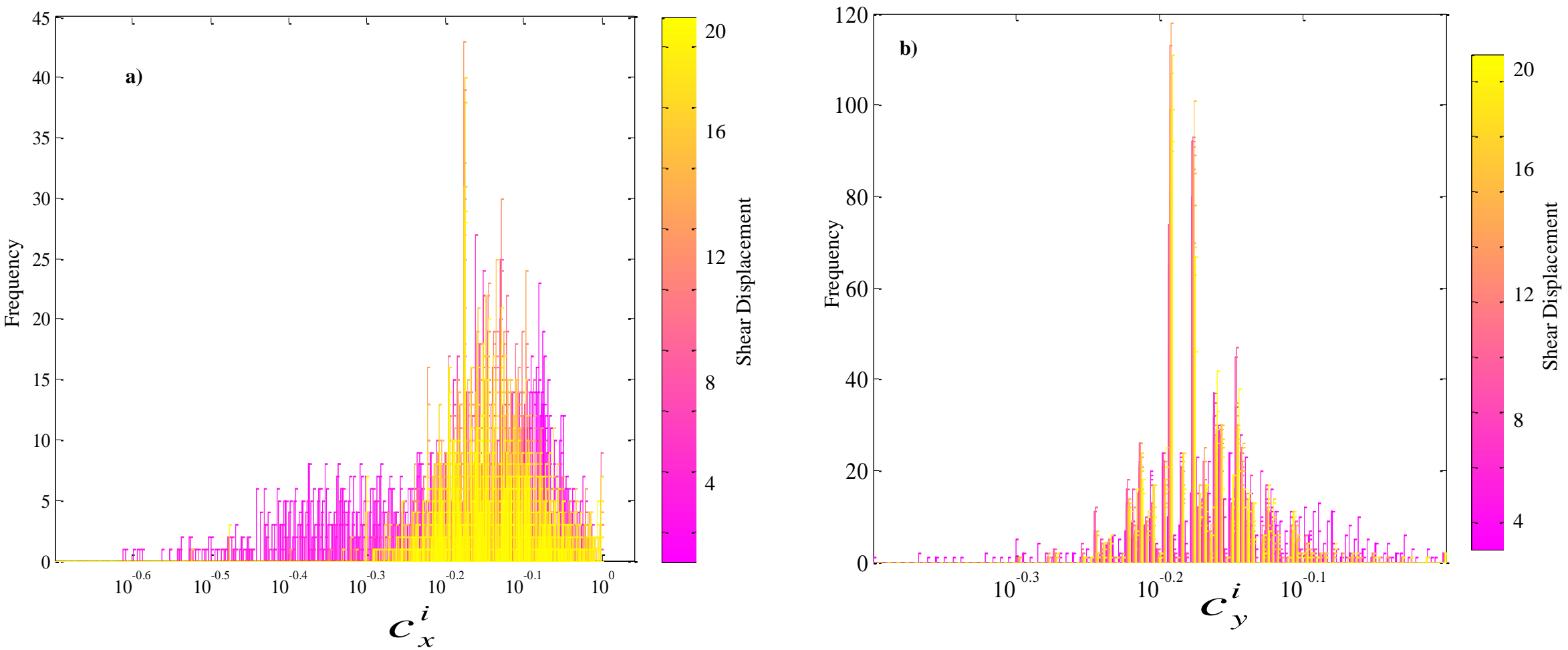

**Figure 7**. Evolution of clustering coefficients frequency (calculated based on the obtained networks- $f(c^{i,j}_{x,y})$) : a) $c_x^{i}$ and b) $c_y^{j}$ during shear displacements (in mm).

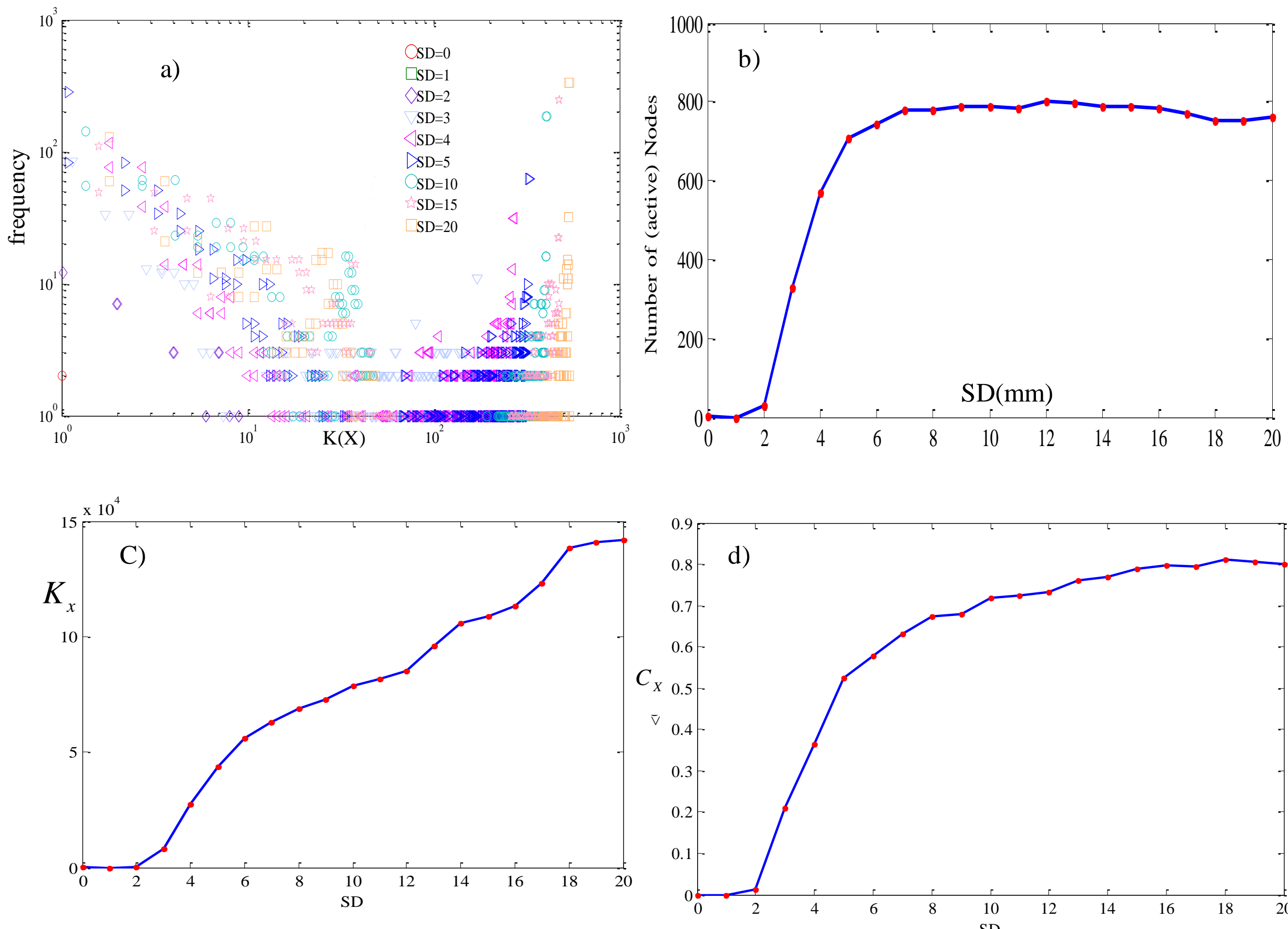

**Figure 8.** The calculated results on the contact profiles network: a) Frequency of nodes connectivity evolution over the shear displacements on X-profiles (perpendicular to shear direction), b) evolution of active nodes, c) Growth of number of edges at X-profiles networks and d) Clustering coefficient versus shear displacement.

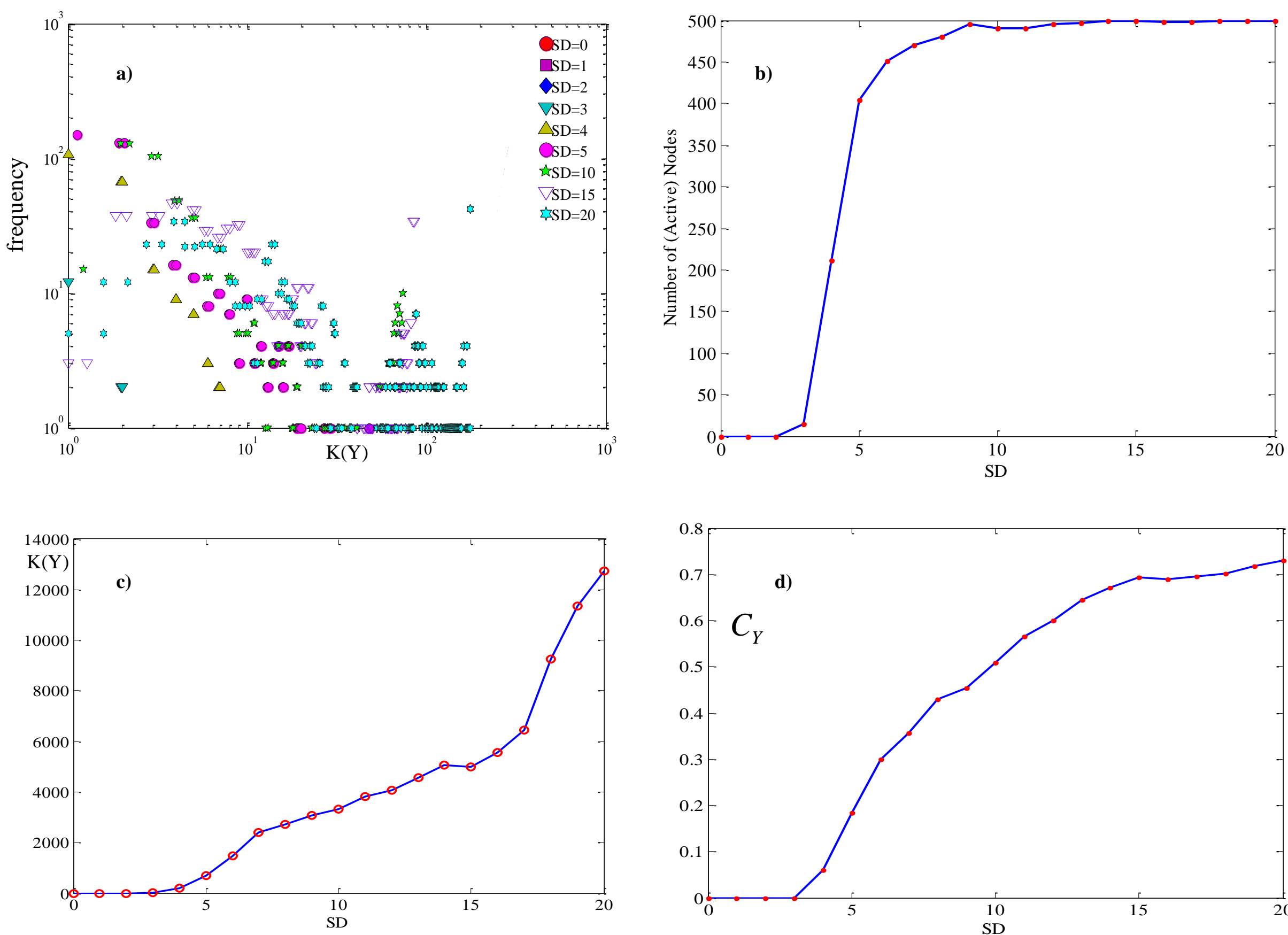

**Figure 9**. Measurements on the contact profiles-network: a) Frequency of nodes connectivity evolution over the shear displacements (SD) on Y-profiles, b) Number of active nodes, c) Growth of number of edges at Y-profiles networks and d) Clustering coefficient versus shear displacement.